\documentclass[11pt,a4paper]{article}

\usepackage[T1]{fontenc}
\usepackage[utf8]{inputenc}
\usepackage{microtype}

\usepackage[margin=2.5cm]{geometry}

\usepackage{amsmath,amssymb}

\usepackage{booktabs}
\usepackage{tabularx}
\usepackage{multirow}
\usepackage{array}

\usepackage{graphicx}
\usepackage{float}
\graphicspath{{figures/}}

\usepackage[dvipsnames]{xcolor}

\usepackage[colorlinks=true,
            linkcolor=NavyBlue,
            citecolor=NavyBlue,
            urlcolor=NavyBlue]{hyperref}

\usepackage{siunitx}
\DeclareSIUnit\kwh{kWh}

\usepackage{titlesec}
\titleformat{\section}{\large\bfseries}{\thesection.}{0.5em}{}
\titleformat{\subsection}{\normalsize\bfseries}{\thesubsection}{0.5em}{}

\usepackage{abstract}

\usepackage{setspace}
\usepackage[numbers,sort&compress]{natbib}
\usepackage{caption}
\usepackage{footmisc}

\newcommand{\coeq}{CO\textsubscript{2}eq}
\newcommand{\htco}{H100 GPU-hour}

\begin{document}

\title{\textbf{Life Cycle Assessment of Pre-training the Lucie 7B
Open-Source Large Language Model on the Jean Zay Supercomputer}}

\author{
  Marc Léobet\textsuperscript{1} \and
  Pierre-François Lavallée\textsuperscript{2} \and
  Jean-Pierre Lorré\textsuperscript{3}
  \\[0.6em]
  \textsuperscript{1}\textit{Mens Data} \quad
  \textsuperscript{2}\textit{CNRS / IDRIS} \quad
  \textsuperscript{3}\textit{LINAGORA}
  \\[0.4em]
  \small From \textit{OpenLLM-France Environmental Report -- v1} -- January 2026
}

\date{}
\maketitle

\begin{abstract}
The environmental impact of training large language models (LLMs) is
increasingly scrutinised, yet most published estimates focus on operational
energy and disclose little about manufacturing (embodied) emissions, water
consumption, or the underlying high-performance computing (HPC)
infrastructure. We present a life cycle assessment (LCA) of the pre-training
of Lucie~7B, an open-source multilingual Foundation Model developed by the
OpenLLM-France consortium and trained on the NVIDIA H100 partition of the
Jean~Zay supercomputer operated by IDRIS (CNRS). The assessment is framed
by the AFNOR SPEC 2314 ``Frugal AI'' reference and applies the Labos
1point5 methodology for greenhouse gas (GHG) accounting in computing. The
study scope extends from data preparation to model validation, and
integrates the full life cycle of the hardware infrastructure: manufacturing
(including raw-material extraction), use (compute, temporary storage, system
administration, cooling), and end-of-life.

We report \textit{(i)}~an annual footprint of \SI{417.5}{\tonne}\,\coeq{}
for the Jean~Zay H100 partition, split almost equally between manufacturing
and operation; \textit{(ii)}~an effective intensity of
\SI{36.7}{\gram}\,\coeq{} per \htco{}; \textit{(iii)}~a total training
footprint of \SI{21}{\tonne}\,\coeq{} for Lucie~7B (574\,564 H100
GPU-hours), inclusive of amortised hardware manufacturing; \textit{(iv)}~on-site
water consumption of approximately \SI{76}{\cubic\metre} for the training
campaign and an annual Water Usage Effectiveness (WUE) of
\SI{0.07}{\litre\per\kwh} for IDRIS; \textit{(v)}~a heat-reuse
factor (ERF) of 0.37 thanks to waste-heat recovery into the urban heating
network. The study contributes one of the few publicly documented LCAs of
an LLM training campaign that explicitly couples operational data with
embodied emissions decomposed by subsystem (compute, storage, power chain,
cooling), and discusses the implications for the design of
frugal-by-construction AI systems in Europe.
\end{abstract}


\vspace{1em}
\hrule
\vspace{1em}

\section{Introduction}

The rapid scaling of deep learning models, and in particular of large
language models (LLMs), has placed their environmental footprint at the
center of academic, industrial, and regulatory debate~\cite{strubell2019,
patterson2021,patterson2022}. Strubell~et~al.~\cite{strubell2019} first
popularised the question for natural language processing, motivating a
stream of work on operational energy use and \coeq{} emissions of
training~\cite{patterson2021,patterson2022,touvron2023a}. Subsequent
contributions have extended the analytical perimeter to
inference~\cite{luccioni2024power}, data-centre water
use~\cite{li2023thirsty,li2025thirsty}, embodied (manufacturing) emissions
of computing hardware~\cite{gupta2022chasing,ligozat2022,wu2022sustainable},
and end-to-end carbon modelling of model training and
serving~\cite{faiz2024llmcarbon}. In parallel, methodological frameworks
specifically targeting digital and AI systems have been proposed at the
European level --- most notably the AFNOR SPEC~2314 ``General Reference
Framework for Frugal AI''~\cite{afnor2314} --- and tools such as the
Labos~1point5 GHG estimator for computing~\cite{labos1p5} have provided
reproducible per-CPU/per-GPU emission factors grounded in actual French
HPC operations.

Despite this growing literature, publicly documented life cycle assessments
(LCAs) of LLM training campaigns remain scarce. The most complete prior
study is the work of Luccioni~et~al.\ on BLOOM-176B~\cite{luccioni2023bloom},
which combines operational measurements with an explicit, if partial,
accounting of hardware manufacturing. Patterson~et~al.~\cite{patterson2021,
patterson2022} disclose detailed operational figures for Google models (T5,
GShard, GPT-3, Switch Transformer) but exclude embodied emissions.
Touvron~et~al.~\cite{touvron2023a} disclose energy and operational \coeq{}
for the LLaMA family, again restricted to operations.
Faiz~et~al.~\cite{faiz2024llmcarbon} propose an end-to-end parametric
model (LLMCarbon) that includes embodied emissions through the ACT
framework~\cite{gupta2022chasing}, but rely on modelled rather than
measured infrastructure data. To our knowledge, no public LCA of an LLM
training campaign has so far combined (a)~measured operational data from a
fully identified European HPC system, (b)~explicit decomposition of embodied
carbon by subsystem (compute, storage, power chain, cooling), (c)~water
use, and (d)~the multi-criteria framing imposed by the AFNOR SPEC
2314~\cite{afnor2314} reference.

This paper addresses that gap. We report an LCA of the pre-training of
Lucie~7B, an open-source Foundation Model released by the OpenLLM-France
consortium~\cite{lucie7b2025}, trained on the NVIDIA H100 partition of the
Jean~Zay supercomputer hosted by IDRIS (CNRS). The analysis follows the
AFNOR SPEC 2314 reference~\cite{afnor2314} and the Labos 1point5
methodology~\cite{labos1p5}, extended to include manufacturing and end-of-life
of the relevant hardware. We restrict the scope to the
data-preparation-through-validation phases of the AFNOR SPEC 2314 life
cycle, and explicitly defer the analysis of inference and downstream
educational services to a forthcoming companion report.

The contributions of this work are threefold. First, we provide reproducible
figures for the annual environmental footprint of an identified European HPC
partition (Jean~Zay H100), with a per-GPU-hour intensity that integrates
manufacturing amortisation. Second, we apply this footprint to the
documented training campaign of an open-source 7-billion-parameter LLM,
yielding a fully traceable training-campaign footprint that includes
embodied carbon. Third, we discuss the implications of two infrastructural
choices --- Direct Liquid Cooling at warm-water regime, and waste-heat
recovery into the urban heating network --- that materially affect both
water and energy balances and that are systematically under-reported in the
LLM literature.

The remainder of the paper is organised as follows. Section~\ref{sec:method}
details the methodology and scope. Section~\ref{sec:case} describes the
case study (Lucie~7B and the Jean~Zay H100 partition).
Section~\ref{sec:results} presents the LCA results: embodied and operational
carbon, annual infrastructure footprint, training-campaign footprint, water
use, and ancillary indicators. Section~\ref{sec:discussion} discusses the
embodied-vs-operational split, the role of waste-heat recovery, and the
comparability of our results with the prior literature.
Section~\ref{sec:limitations} enumerates the limitations of this v1 report
and the work programme for a v2.

\section{Methodology}
\label{sec:method}

\subsection{Frameworks}

The assessment follows the AFNOR SPEC 2314:2024 reference for frugal
AI~\cite{afnor2314}, which structures the life cycle of an AI system
(``SIA'') into seven phases --- data, design, training, validation,
deployment, use, and end-of-life --- and prescribes a multi-criteria scope.
For greenhouse gas accounting at the HPC infrastructure layer, we apply the
Labos 1point5 methodology~\cite{labos1p5}, which provides per-CPU and
per-GPU emission factors derived from a bottom-up analysis of actual French
HPC centres~\cite{genci2020}. We rely on the methodology rather than on the
published 2020 figures, because Labos 1point5 figures predate the H100
generation; the IDRIS team performed an updated computation for this study
using the same methodology, applied to the 2024--2025 Jean~Zay H100
configuration. The framework is consistent in spirit with ISO
14040/14044~\cite{iso14040}, although a full ISO-compliant critical review
has not been conducted at this stage.

\subsection{Goal and scope}

The goal of the study is to quantify the environmental impact of the
pre-training of Lucie~7B, including the share attributable to the
manufacturing of the underlying HPC infrastructure. The functional scope
spans the AFNOR SPEC 2314 phases from data acquisition to model validation.
Inference and downstream services are excluded from this v1 report and will
be addressed in v2 once empirical inference workloads are available.

As a consequence of the open-source nature of the model --- released under
terms compatible with the Open Source Initiative AI
definition~\cite{osiai2024} --- the SIA can be re-used and modified by
unknown third parties, which structurally limits the analytical reach of any
single LCA. We therefore restrict the analysis to the impacts under the
responsibility of the OpenLLM-France consortium during pre-training and
validation.

\subsection{System boundaries}

Following the Arcep/Ademe taxonomy of digital footprint~\cite{arcep2022},
the analysis focuses on the data-centre tier ($\approx$16\,\% of digital
GHG emissions in France in 2020), and acknowledges that the terminal
($\approx$79\,\%) and network ($\approx$5\,\%) tiers fall outside the scope
of training. Within the data-centre tier, the boundary includes: compute
nodes (CPUs and GPUs), temporary storage (SSD and HDD), system
administration of compute and storage platforms, the secured power chain
(UPS and distribution), and cooling (servers and UPSs). It excludes:
building construction and maintenance (amortised as over 50 years old),
maintenance labour, and personnel activity (technical and administrative),
all due to lack of data. Transport of equipment is considered negligible.
Following Labos 1point5~\cite{labos1p5}, waste-heat recovery is excluded
from the GHG perimeter; we report it separately as an ancillary indicator
(Section~\ref{sec:ere}).

\subsection{Functional units}

Two functional units (FUs) are used. The \emph{infrastructure-level} FU is
one hour of intensive compute on one NVIDIA H100 GPU at IDRIS, including
the proportional share of storage, power, cooling, and embodied emissions.
The \emph{campaign-level} FU is the full pre-training of Lucie~7B, defined
by the GPU-hour budget consumed during the training campaign on Jean~Zay.
Comparisons with other models are reported in Section~\ref{sec:comparability}
with explicit reservations: the absence of a normalisation by model
performance and by training tokens limits the strict comparability across
campaigns.

\subsection{Environmental indicators}

The AFNOR SPEC 2314 reference prescribes two high-priority and four
medium-priority indicators. We retain: climate change (kg \coeq{}); abiotic
resource depletion --- minerals and metals (kg Sb\,eq); water consumption
and withdrawal (\si{\cubic\metre}); and we report energy consumption (Wh)
as a primary auxiliary indicator. The remaining medium-priority indicators
(ocean acidification, fine-particulate emissions, ionising radiation) are
qualitatively assessed and considered marginal for the scope of this study;
their full quantification is left for v2. This scope reduction is a known
limitation discussed in Section~\ref{sec:limitations}.

\subsection{Allocation}

Allocation between research projects sharing the Jean~Zay H100 partition is
performed on a GPU-hour basis. Lucie~7B consumed 574\,564 H100 GPU-hours,
allocated against an effective annual partition capacity
(cf.\ Section~\ref{sec:partition}). The allocation is transparent for
compute, and propagated to storage, power, and cooling proportionally to the
partition share. Storage hardware is allocated linearly to the partition by
capacity.

\subsection{Equipment lifetimes and amortisation}

Manufacturing emissions are amortised over technical lifetimes that reflect
actual IDRIS practice. GPUs and compute nodes are used in intensive
computing for 5 to 6 years (85\,\%--94\,\% of time at Jean~Zay) and then
re-used in less intensive configurations for several additional years. We
adopt a 10-year amortisation window for compute (6 years intensive + 4
years extended use), 9 years for storage (with 50\,\% effective extension
via re-conditioning, as observed for the N1 storage of the previous
Jean~Zay extension), 25 years for the power chain, and 20 years for the
cooling system. These choices are documented in Section~\ref{sec:embodied}.

\section{Case Study Setup}
\label{sec:case}

\subsection{The Lucie 7B model}

Lucie~7B is a 7-billion-parameter open-source Foundation Model developed by
the OpenLLM-France consortium~\cite{lucie7b2025}. It is distributed under
terms aligned with the Open Source Initiative AI definition~\cite{osiai2024},
and is qualified as a general-purpose AI system in the sense of Article~3-63
of the EU AI Act~\cite{aiact2024}, without falling into the systemic-risk
category. Pre-training was conducted on a multilingual corpus, with
substantial coverage of French and other European languages (English,
German, Spanish and Italian).

\subsection{The Jean Zay H100 partition}
\label{sec:partition}

The Jean~Zay supercomputer is operated by IDRIS, the major CNRS centre for
high-performance scientific computing, and is part of the GENCI national HPC
ecosystem (with TGCC and CINES). The fourth Jean~Zay extension --- installed
in 2024, officially inaugurated on 13~May 2025 --- adds a partition based on
Eviden BullSequana XH3000 racks, comprising 14~racks, 364~dual-socket
compute nodes (Intel Sapphire Rapids 48-core, 512\,GB RAM), and 1\,456
NVIDIA H100 80\,GB SXM5 GPUs (4 GPUs per node). The partition reaches
125.9\,PFLOP/s peak and provides up to 12.75 million H100 GPU-hours per
year to the scientific community. Cooling is provided by a Direct Liquid
Cooling (DLC) loop with \SI{30}{\celsius}/\SI{36}{\celsius}
inlet/outlet warm-water regime; waste heat is partially recovered for urban
heating.

\subsection{Training campaign}

The Lucie~7B pre-training campaign ran for approximately three months on the
Jean~Zay H100 partition and consumed 574\,564 H100 GPU-hours, corresponding
to roughly 5\,\% of the partition annual budget. Tokenisation of the
training corpus ($\approx$2.3 trillion tokens after multiple passes) was
performed on Jean~Zay CPU resources prior to the GPU-hosted pre-training.

\section{Results}
\label{sec:results}

\subsection{Embodied (manufacturing) emissions}
\label{sec:embodied}

Manufacturing emissions are estimated bottom-up by subsystem, drawing on
Labos~1point5 figures~\cite{labos1p5,genci2020} for compute, storage,
power, and cooling, and on the NVIDIA H100 product carbon
footprint~\cite{nvidiah100pcf} for the GPU dies and modules. The latter is
acknowledged to follow a partial scope (Section~\ref{sec:limitations}). The
corresponding totals for the H100 partition are shown in
Table~\ref{tab:embodied_total}.

\begin{table}[ht]
\centering
\caption{Total manufacturing emissions of the Jean~Zay H100 partition by
subsystem, with amortisation lifetimes.}
\label{tab:embodied_total}
\begin{tabularx}{\linewidth}{Xcc}
\toprule
\textbf{Subsystem} & \textbf{Manuf.\ (t\,\coeq{})} &
  \textbf{Lifetime} \\
\midrule
Compute (364 dual-CPU nodes + 1\,456 H100 GPUs) & 748 & 10 years \\
Storage (SSD 7.6\,PB + HDD 39\,PB)             & 541 & 9 years  \\
Power chain                                      & 268 & 25 years \\
Cooling system (warm-water DLC)                  & 612 & 20 years \\
\midrule
\textbf{Total (capital)}                         & \textbf{2\,169} & --- \\
\bottomrule
\end{tabularx}
\end{table}

Per-node compute manufacturing is decomposed as \SI{1400}{\kilo\gram}\,\coeq{}
for the dual-CPU host node and $4 \times \SI{164}{\kilo\gram}$\,\coeq{} for
the H100 GPUs, totalling \SI{2056}{\kilo\gram}\,\coeq{} per node. For
storage, the 2020 Jean~Zay reference values are linearly extrapolated to
the current configuration (7.6\,PB SSD across N1 and N2+ tiers, 39\,PB
HDD), yielding \SI{430}{\tonne}\,\coeq{} for SSD (servers + media) and
\SI{111}{\tonne}\,\coeq{} for HDD. The power chain and cooling system are
unchanged from the 2020 baseline and their values are reused.

Two flow-related sources are accounted for at the partition level:
refrigerant fugitive emissions (R134a,
$\SI{14}{\kilo\gram}\text{/year} \times \SI{1430}{\kilo\gram}\,\coeq{}/\text{kg}
= \SI{15}{\tonne}\,\coeq{}\text{/year}$) and emergency diesel consumption
($\approx\SI{900}{\litre}\text{/year}$, \SI{2.8}{\tonne}\,\coeq{}/year).
Replacement of R134a by a fluid compliant with the ``bringing a substantial
contribution to climate change mitigation'' classification of EU Taxonomy on
sustainable investment~\cite{eutaxonomy2021} is planned within 18~months of
this report.

Annualised manufacturing emissions for the H100 partition, including fluid
leakage and diesel, are detailed in Table~\ref{tab:embodied_annual}.

\begin{table}[ht]
\centering
\caption{Annualised manufacturing emissions of the Jean~Zay H100 partition.}
\label{tab:embodied_annual}
\begin{tabular}{lll}
\toprule
\textbf{Item} & \textbf{Annualised (t\,\coeq{}/year)} & \textbf{Comment} \\
\midrule
Compute (10\,y)         & 74.8  & with GPU/memory re-use \\
Storage (9\,y)          & 60.1  & with N1 re-conditioning \\
Power chain (25\,y)     & 10.7  & \\
Cooling (20\,y)         & 30.6  & \\
Refrigerant + diesel    & 18.0  & yearly flows, no amortisation \\
\midrule
\textbf{Total annualised} & \textbf{194.2} & \\
\bottomrule
\end{tabular}
\end{table}

\subsection{Operational emissions}

Operational emissions are computed from monitored electricity consumption
of the H100 partition at IDRIS over the last 273~days of 2025, scaled to a
yearly basis. Total Jean~Zay electricity consumption is
\SI{21864}{\mega\watt\hour}/year, decomposed as: compute partitions
\SI{15127}{\mega\watt\hour}, storage infrastructure \SI{935}{\mega\watt\hour},
technical infrastructure (power conditioning and cooling)
\SI{5803}{\mega\watt\hour}. The H100 partition specifically draws
\SI{8512}{\mega\watt\hour}/year for compute and storage; the
proportional share of technical infrastructure is
\SI{1787}{\mega\watt\hour}/year (PUE\,=\,1.21), bringing the total
H100 partition electricity to \SI{10299}{\mega\watt\hour}/year.

With a French electricity emission factor of
\SI{21.7}{\gram\,\coeq{}\per\kwh} in 2024~\cite{rte2024},
operational emissions of the H100 partition reach
\SI{223.5}{\tonne}\,\coeq{}/year. The 2025 emission factor is lower
(\SI{19.6}{\gram\,\coeq{}\per\kwh} according to RTE), which
will mechanically reduce the operational figure for following campaigns of
the project.

\subsection{Annual infrastructure footprint and per-GPU-hour intensity}

Aggregating manufacturing and operational emissions, the total annual carbon
footprint of the Jean~Zay H100 partition is:

\begin{equation}
  194.2 + 223.5 = \SI{417.5}{\tonne}\,\coeq{}\,/\,\text{year}
\end{equation}

with manufacturing and operations contributing nearly equal shares (46\,\%
and 54\,\% respectively). Applying the Labos~1point5 methodology, this
total is divided by the effective annual GPU-hour budget of the partition
($\approx$11.39~million GPU-hours, corresponding to $\approx$88\,\% average
utilisation of the 12.75~million nominal hours) to yield a partition-level
intensity of:

\begin{equation}
  \boxed{\SI{36.7}{\gram}\,\coeq{}\text{ per H100 GPU-hour (LCA-inclusive)}}
\end{equation}

\begin{figure}[ht]
  \centering
  \includegraphics[width=\linewidth]{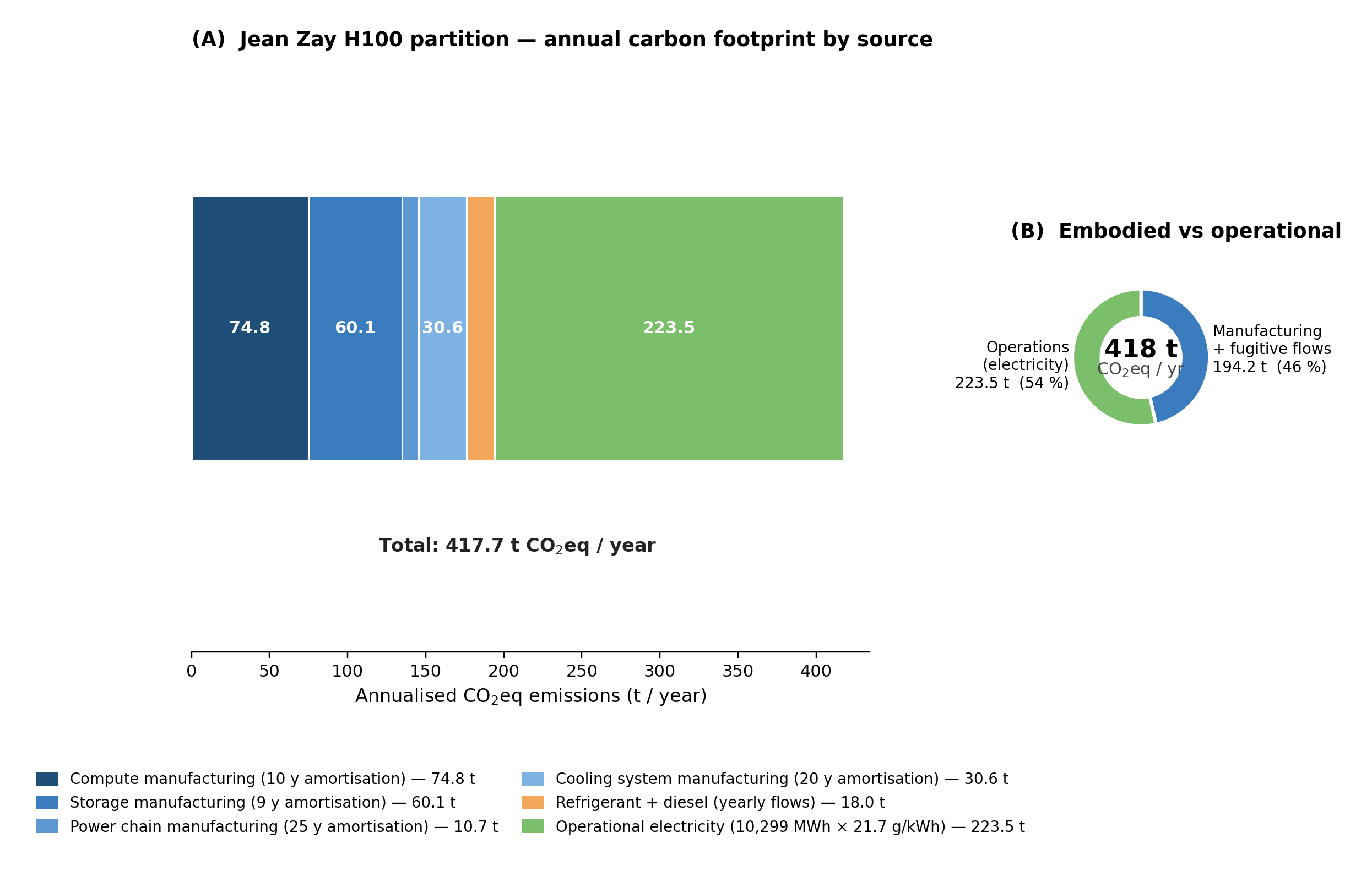}
  \caption{\textbf{Annual environmental footprint of the Jean~Zay H100
    partition (CNRS/IDRIS).} Panel~(A): decomposition by source --- four
    manufacturing components (compute, storage, power chain, cooling)
    amortised over their respective technical lifetimes, yearly fugitive
    flows (refrigerant + diesel), and operational electricity. Panel~(B):
    aggregate split between manufacturing/fugitive flows and operations.
    Lucie~7B (574\,564 GPU-h, $\approx$5\,\% of the partition annual
    budget) is allocated \SI{21.1}{\tonne}\,\coeq{} accordingly.}
  \label{fig:footprint}
\end{figure}

Figure~\ref{fig:footprint} visualises the decomposition.

\subsection{Lucie 7B training-campaign footprint}

Multiplying the per-GPU-hour intensity by the 574\,564 H100 GPU-hours
consumed by the Lucie~7B campaign yields a training-campaign footprint of:

\begin{equation}
  \boxed{\SI{21.1}{\tonne}\,\coeq{}\text{ for the Lucie 7B pre-training
  (LCA-inclusive)}}
\end{equation}

This figure includes the amortised share of manufacturing emissions of the
entire H100 partition allocated to the campaign, in addition to operational
electricity. Tokenisation of the training corpus is treated separately
(Section~\ref{sec:dataprep}) and contributes negligibly to the total.

\subsection{Water consumption}

Water at IDRIS is used for adiabatic cooling of the warm-water DLC loop and
for refrigeration of ancillary equipment. Total annual withdrawal of
municipal water is \SI{3000}{\cubic\metre}, of which \SI{1500}{\cubic\metre}
are consumed by evaporation; the remainder is returned to the wastewater
network after blowdown. The annual Water Usage Effectiveness (WUE) is:

\begin{equation}
  \mathrm{WUE} = \frac{1.5 \times 10^6\,\si{\litre}}
                      {21\,864 \times 10^3\,\si{\kwh}}
               = \SI{0.07}{\litre\per\kwh}
\end{equation}

This compares favourably with the 0.25--\SI{0.30}{\litre\per\kwh}
range reported in the 2025 ARCEP survey of French data centres~\cite{arcep2025}.
IDRIS is not located in a hydric-stress zone.

Allocating to Lucie~7B at the 5\,\% share of partition GPU-hours yields:

\begin{equation}
  \boxed{\approx\SI{76}{\cubic\metre}\text{ of on-site water consumption
  attributable to Lucie 7B pre-training}}
\end{equation}

As a complementary indicator, the off-site water intensity associated with
French electricity production (Energy Water Intensity Factor, EWIF) is
reported by EDF at \SI{0.86}{\litre\per\kwh} in
2024~\cite{edf2024}. Off-site water has not been added to the headline
figure in this v1; its inclusion is discussed in Section~\ref{sec:limitations}
and planned for v2 to comply with the dual on-site/off-site accounting
framework proposed by Li~et~al.~\cite{li2023thirsty,li2025thirsty}.

\subsection{Data preparation}
\label{sec:dataprep}

Tokenisation of 120~billion tokens required 20~hours on 40~CPU cores;
processing the full 2.3-trillion-token training corpus thus required
383~hours on 40~CPU cores, i.e.\ 15\,333~CPU-hours. With an emission factor
of \SI{1.1}{\gram\,\coeq{}\per\text{CPU-core-hour}}~\cite{labos1p5}, this
phase emits $\approx$\SI{15}{\kilo\gram}\,\coeq{}, which is negligible
compared with the GPU-hosted pre-training. Tokeniser training itself is
sub-marginal (a few CPU-hours).

\subsection{Ancillary indicators (ERF, REF, ERE)}
\label{sec:ere}

Three additional indicators, formalised by The Green Grid~\cite{greengrid2010}
and partly recognised by EU Regulation~2024/1364~\cite{eu2024dc}, complement
the picture. The \emph{Energy Reuse Factor} (ERF) --- ratio of useful waste
heat re-used to total energy consumed --- is 0.37 at Jean~Zay; equivalently,
$\approx$\SI{6500}{\mega\watt\hour}/year of useful heat is recovered
into a district heating network, equivalent to the heating demand of
$\approx$1\,500 newly built dwellings. In winter operation, up to 80\,\% of
the partition energy is reused as heat. The \emph{Renewable Energy Factor}
(REF) inherits the French electricity mix at 27.8\,\% renewable in 2024.
The \emph{Energy Reuse Effectiveness} (ERE), a heat-recovery-aware analogue
of PUE, is 0.93 for the H100 partition.

\section{Discussion}
\label{sec:discussion}

\subsection{Embodied/operational split}

A central finding of this study is that, on a low-carbon electricity grid
such as the French one (\SI{21.7}{\gram\,\coeq{}\per\kwh}), the
embodied (manufacturing) component is not dominated by operations: at the
partition level, the split is approximately 46/54. This contrasts with
results obtained on higher-carbon grids --- where operations typically dwarf
embodied --- and is consistent with the qualitative position taken by
Wu~et~al.~\cite{wu2022sustainable} and Gupta~et~al.~\cite{gupta2022chasing}
that embodied carbon becomes the bottleneck on clean grids. For Jean~Zay
specifically, the implication is that further mitigation will increasingly
depend on hardware-side levers (lifetime extension, re-use, lower-carbon GPU
manufacturing) rather than electricity decarbonisation alone.

\subsection{Comparability with prior literature}
\label{sec:comparability}

Direct numerical comparison with prior published LLM training footprints
requires care. Reported LLaMA-2 7B operational figures from
Touvron~et~al.~\cite{touvron2023b} ($\approx$\SI{31}{\tonne}\,\coeq{})
reflect operations only, on a different grid (US-based), with A100 rather
than H100 GPUs, and across a different number of training tokens; they are
not framed as functional units in the LCA sense. The
\SI{21.1}{\tonne}\,\coeq{} we report for Lucie~7B is LCA-inclusive
(operations + amortised manufacturing) but on a low-carbon grid, with
newer-generation GPUs, and again with a distinct training corpus and target
performance. We therefore refrain from headline-level efficiency claims and
limit the comparison to a contextual data point. A full normalised
comparison --- per training token, per FLOP, or per benchmark performance
unit --- is left for v2 and would benefit from coordinated disclosure across
LLM developers.

\subsection{Heat recovery as a systemic differentiator}

The ERF of 0.37 achieved by Jean~Zay translates into approximately
\SI{6500}{\mega\watt\hour}/year of useful heat injected into the urban
heating system. Although excluded from the GHG perimeter under the
Labos~1point5 convention, this avoided heat displaces $\approx$1\,500
dwellings worth of heating energy, with a non-trivial system-level carbon
benefit. We argue that ERF and ERE deserve to be reported alongside PUE and
WUE in any LCA of HPC-hosted training campaigns, since they capture an
integration with surrounding infrastructure that hyperscale-only metrics
miss.

\subsection{Frugal-by-construction infrastructure}

Two infrastructural choices materially shape the results: warm-water Direct
Liquid Cooling (\SI{30}{\celsius} inlet) and waste-heat recovery. The DLC
regime drives the favourable WUE (\SI{0.07}{\litre\per\kwh}) and
limits the cooling power overhead (PUE\,=\,1.21), while heat recovery turns
part of the operational waste into useful output. Both choices were design
decisions of the GENCI/IDRIS infrastructure long predating the Lucie~7B
campaign, and they illustrate the importance of accounting for
infrastructure-level decisions when reasoning about model-level frugality.
The AFNOR SPEC 2314 reference~\cite{afnor2314} is explicit on this point
and provides a useful structuring of efficiency vs.\ frugality of services
delivered.

\section{Limitations and Future Work}
\label{sec:limitations}

This is a v1 report and we list its limitations transparently.

\paragraph{Inference and downstream services.} The study covers only the
AFNOR SPEC 2314 phases up to model validation; inference and downstream
educational services (planned within OpenLLM-France) are excluded and will
be addressed in v2 once the per-prompt processing time and serving
infrastructure are stabilised.

\paragraph{Indicator scope.} The indicator scope is reduced to climate
change, water, and (qualitatively) abiotic resources for the v1; ocean
acidification, fine-particulate emissions, and ionising radiation are
flagged as marginal but not formally quantified. A v2 should at minimum
quantify abiotic resource depletion (kg\,Sb\,eq) for compute and storage
subsystems, drawing on inventories such as Ligozat~et~al.~\cite{ligozat2022}
and Wu~et~al.~\cite{wu2022sustainable}.

\paragraph{Off-site water footprint.} The water footprint reported here is
restricted to on-site cooling. The off-site EWIF associated with electricity
generation ($\approx$\SI{0.86}{\litre\per\kwh} for the French
mix in 2024~\cite{edf2024}) is documented but not added to the headline
figure. The dual on-site/off-site accounting proposed by
Li~et~al.~\cite{li2023thirsty,li2025thirsty} should be implemented in v2.

\paragraph{GPU embodied emissions.} The embodied figure for the H100 GPU
(\SI{164}{\kilo\gram}\,\coeq{}) is taken from the NVIDIA product carbon
footprint~\cite{nvidiah100pcf}, whose scope is acknowledged to be partial.
Triangulation against the ACT model~\cite{gupta2022chasing} and bottom-up
estimates from foundry-level inventories should be included in v2 with an
explicit sensitivity analysis.

\paragraph{Amortisation sensitivity.} The amortisation windows (10\,y for
compute, 9\,y for storage, 25\,y for the power chain, 20\,y for cooling)
reflect IDRIS practice but are admittedly point estimates; a sensitivity
analysis with ranges of $\pm$2--3~years is straightforward and should be
reported. Likewise, the Labos~1point5 methodology suggests an overall
$\pm$20\,\% uncertainty, which we have not formally propagated.

\paragraph{Failed runs and exploration.} The GPU-hour figure used for
Lucie~7B (574\,564\,h) corresponds to the documented training campaign;
failed runs and hyperparameter exploration prior to the final campaign are
not separately accounted for. In line with practice in the field, this
likely under-estimates the true campaign-wide footprint by an unknown
factor; we encourage future work in the community to disclose run-level
instrumentation logs.

\paragraph{ISO critical review.} While we apply the AFNOR SPEC 2314
reference, we have not commissioned the formal third-party critical review
procedure foreseen by ISO~14071. The framework alignment is therefore
methodological rather than strictly ISO-certified.

\section{Conclusion}

We have presented a v1 life cycle assessment of the pre-training of
Lucie~7B, an open-source Foundation Model trained on the NVIDIA H100
partition of the Jean~Zay supercomputer. The study quantifies an annual
partition footprint of \SI{417.5}{\tonne}\,\coeq{}, an LCA-inclusive
intensity of \SI{36.7}{\gram}\,\coeq{} per H100 GPU-hour, and a
training-campaign footprint of \SI{21.1}{\tonne}\,\coeq{} for Lucie~7B,
with associated on-site water consumption of $\approx$\SI{76}{\cubic\metre}.
The near-equal split between embodied and operational emissions on the French
grid argues for an increased focus on hardware-side mitigation levers in
low-carbon HPC contexts. Two design decisions of the underlying
infrastructure --- warm-water DLC and waste-heat recovery into urban heating
--- make a measurable difference to both water and useful-energy balances,
and we suggest that ERF and ERE be reported as standard indicators alongside
PUE and WUE in LLM environmental disclosures. The limitations enumerated in
Section~\ref{sec:limitations} will be addressed in a v2 report integrating
inference, sensitivity analysis, dual on-site/off-site water accounting, and
broader indicator coverage.

\section*{Acknowledgements}

The authors thank the IDRIS technical and scientific staff for the
operational data on the Jean~Zay H100 partition, the Labos~1point5
collective for the methodological foundations of HPC GHG accounting, and
the OpenLLM-France consortium members for their contribution to the design
and training of the Lucie~7B model. The training campaign was performed
using HPC resources from GENCI--IDRIS (Grant 2024-GC011015444).

This work was supported by Bpifrance, as part of the OpenLLM-France project.

\section*{Licence}

This report is released under the Creative Commons
Attribution-ShareAlike 4.0 International (CC~BY-SA~4.0) licence.

\bibliography{references}

\end{document}